\documentclass{revtex4}
\def\[{\left\lbrack}
\def\]{\right\rbrack}

\def\({\left(}
\def\){\right)}
\newcommand{\be}{\begin{equation}}
\newcommand{\ee}{\end{equation}}
\newcommand{\ea}{\end{eqnarray}}
\newcommand{\ba}{\begin{eqnarray}}

\begin{document}

\title{Hamiltonian symplectic embedding of the massive noncommutative $U(1)$ Theory}

\author{C. Neves, W. Oliveira}
\email{cneves, wilson@fisica.ufjf.br}
\affiliation{Departamento de F\'{\i}sica, ICE, Universidade Federal de Juiz de Fora,\\
36036-330, Juiz de Fora, MG, Brasil}
\author{D. C. Rodrigues and C. Wotzasek}
\email{clovis, cabral@if.ufrj.br}
\affiliation{Instituto de F\'{\i}sica, Universidade Federal do Rio de Janeiro,\\
21945-970, Rio de Janeiro, RJ, Brasil}

\begin{abstract}
\noindent
We show that the massive noncommutative $U(1)$ theory is embedded in a gauge theory using an alternative systematic way \cite{ANO}, which is based on the symplectic framework. The embedded Hamiltonian density is obtained after a finite number of steps in the iterative symplectic process, oppositely to the result proposed using the BFFT formalism \cite{RB}. This alternative formalism of embedding shows how to get a set of dynamically equivalent embedded Hamiltonian densities.
\end{abstract}

\maketitle


\setlength{\baselineskip} {20 pt}

\section{Introduction}

The embedding mechanism, firstly suggest by Faddeev and Shatashivilli\cite{FS}, has been a successful constraint conversion procedure over the last decades. The main concept behind this procedure resides on the enlargement of the phase space with the introduction of new variables, called Wess-Zumino (WZ) variables, which changes the second class nature of the constraint to first one. This procedure has been explored in different context\cite{BFFT,IJMP,NW,ANO} in order to avoid some problems that affect the quantization process of some theories, such as chiral theory, where the anomaly obstructs the quantization mechanism, and nonlinear models, where the operator ordering ambiguities arise. It is opportune to comment here that the proposed embedding procedure, be it applied to commutative or noncommutative theories, is to unveil the origin of the ambiguities of all embedding approaches. 

The great deal of interest in  noncommutative (NC) field theories  started when it was noted that noncommutativite spaces naturally arise in string theory with a constant background magnetic field in the presence of D-branes. It is opportune to mention here that this noncommutativity in the context of the string theory with a constant background magnetic field in the presence of D-branes was eliminated constructing a mechanical system which reproduces classical dynamics of the string \cite{string}. Besides their origin in strings theories and branes, noncommutative (NC) field theories have been studied extensively in many branches of the Physics \cite{other,RB}. 

In order to obtain the noncommutative version of a field theory one replace the usual product of fields in the action by the Moyal product, defined as

\be
\phi_1 (x)\star \phi_2 (x) =exp\left( {i\over 2} \theta^{\mu \nu} \partial^{x}_{\mu}\partial^{y}_{\nu} \right) \phi_1 (x) \phi_2 (y)\mid_{x=y} \nonumber
\ee
where $\theta^{\mu \nu}$ is a real and antisymmetric constant matrix. As a consequence, NC theories are highly nonlocal. We also note that Moyal product of two fields in the action is the same as the usual product ({\it vide} appendix), provide we discard boundary terms. Thus, the noncommutativity affects just the vertices.

Recently, the embedded version of the massive noncommutative $U(1)$ theory was obtained through the BFFT constraint conversion scheme\cite{RB}. In this work, the authors showed how to obtain a set of second class constraints and Hamiltonian which form an involutive system of dynamical quantities. However, both the constraints and Hamiltonian were expressed as a series of Moyal commutators among the variables belonging to the WZ extended phase space. Our goal in the present work is to propose a embedded version for the massive noncommutative $U(1)$ theory where the embedded Hamiltonian density is not expressed as expansion on the WZ variables but as a finite sum. To this end, we will use the symplectic embedding formalism ({\it vide} Sec. \ref{sec:II}) \cite{ANO}.

Our paper is organized as follows. In Sec. \ref{sec:II}, we present an overview of the symplectic embedding formalism. In Sec. \ref{sec:III}, we analyze the symplectic quantization of the noncommutative massive $U(1)$ theory, and compute the Dirac brackets among the phase space. In Sec. \ref{sec:IV}, we investigate the embedded version for the noncommutative massive $U(1)$ from the symplectic embedding point of view. We note that after a finite number of steps of the iterative symplectic embedding process, we obtain an embedded Hamiltonian density. In consequence, this Hamiltonian density has a finite number of WZ terms, oppositely to \cite{RB}. In Sec. \ref{sec:V}, we present some concluding remarks. In appendix, we list some properties of the Moyal product that we use in this paper.

\section{General formalism}
\label{sec:II}

In this section, we describe the alternative embedding technique that changes the second class nature of a constrained system to the first one. This technique follows the Faddeev and Shatashivilli idea\cite{FS} and is based on a contemporary framework that handles constrained models, namely, the symplectic formalism\cite{FJ,BC}. 

In order to systematize the symplectic embedding formalism, we consider a general noninvariant mechanical model whose dynamics is governed by a Lagrangian ${\cal L}(a_i,\dot a_i,t)$, (with $i=1,2,\dots,N$), where $a_i$ and $\dot a_i$ are the space and velocities variables, respectively. Notice that this model does not result in the loss of generality or physical content. Following the symplectic method the zeroth-iterative first-order Lagrangian 1-form is written as
 
\begin{equation}
\label{2000}
{\cal L}^{(0)}dt = A^{(0)}_\theta d\xi^{(0)\theta} - V^{(0)}(\xi)dt,
\end{equation}
where the symplectic variables are

\be
\xi^{(0)\alpha} =  \left\{ \begin{array}{ll}
                               a_i, & \mbox{with $\alpha=1,2,\dots,N $,} \\
                               p_i, & \mbox{with $\alpha=N + 1,N + 2,\dots,2N ,$}       
                           \end{array}
                     \right.
\ee
$A^{(0)}_\alpha$ are the canonical momenta and $V^{(0)}$ is the symplectic potential. The symplectic tensor is given by

\begin{eqnarray}
\label{2010}
f^{(0)}_{\alpha\beta} = {\partial A^{(0)}_\beta\over \partial \xi^{(0)\alpha}}
-{\partial A^{(0)}_\alpha\over \partial \xi^{(0)\beta}}.
\end{eqnarray}
When the two-form $f \equiv \frac{1}{2}f_{\theta\beta}d\xi^\theta \wedge d\xi^\beta$ is singular, the symplectic matrix (\ref{2010}) has a zero-mode $(\nu^{(0)})$ that generates a new constraint when contracted with the gradient of the symplectic potential, 
\begin{equation}
\label{2020}
\Omega^{(0)} = \nu^{(0)\alpha}\frac{\partial V^{(0)}}{\partial\xi^{(0)\alpha}}.
\end{equation}
This constraint is introduced into the zeroth-iterative Lagrangian 1-form, Eq.(\ref{2000}), through a Lagrange multiplier $\eta$, generating the next one

\begin{eqnarray}
\label{2030}
{\cal L}^{(1)}dt &=& A^{(0)}_\theta d\xi^{(0)\theta} + d\eta\Omega^{(0)}- V^{(0)}(\xi)dt,\nonumber\\
&=& A^{(1)}_\gamma d\xi^{(1)\gamma} - V^{(1)}(\xi)dt,
\end{eqnarray}
with $\gamma=1,2,\dots,(2N + 1)$ and 

\begin{eqnarray}
\label{2040}
V^{(1)}&=&V^{(0)}|_{\Omega^{(0)}= 0},\nonumber\\
\xi^{(1)_\gamma} &=& (\xi^{(0)\alpha},\eta),\\
A^{(1)}_\gamma &=&(A^{(0)}_\alpha, \Omega^{(0)}).\nonumber
\end{eqnarray}
As a consequence, the first-iterative symplectic tensor is computed as

\begin{eqnarray}
\label{2050}
f^{(1)}_{\gamma\beta} = {\partial A^{(1)}_\beta\over \partial \xi^{(1)\gamma}}
-{\partial A^{(1)}_\gamma\over \partial \xi^{(1)\beta}}.
\end{eqnarray}
If this tensor is nonsingular, the iterative process stops and the Dirac's brackets among the phase space variables are obtained from the inverse matrix $(f^{(1)}_{\gamma\beta})^{-1}$ and, consequently, the Hamilton equation of motion can be computed and solved as well \cite{gotay}. It is well known that a physical system can be described in terms of a symplectic manifold $M$, classically at least. From a physical point of view, $M$ is the phase space of the system while a nondegenerate closed 2-form $f$ can be identified as being the Poisson bracket. The dynamics of the system is  determined just specifying a real-valued function (Hamiltonian) $H$ on phase space, {\it i.e.}, one these real-valued function solves the Hamilton equation, namely,
\be
\label{2050a1}
\iota(X)f=dH,
\ee
and the classical dynamical trajectories of the system in phase space are obtained. It is important to mention that if $f$ is nondegenerate, Eq.(\ref{2050a1}) has an unique solution. The nondegeneracy of $f$ means that the linear map $\flat:TM\rightarrow T^*M$ defined by $\flat(X):=\flat(X)f$ is an isomorfism, due to this, the Eq.(\ref{2050a1}) is solved uniquely for any Hamiltonian $(X=\flat^{-1}(dH))$. On the contrary, the tensor has a zero-mode and a new constraint arises, indicating that the iterative process goes on until the symplectic matrix becomes nonsingular or singular. If this matrix is nonsingular, the Dirac's brackets will be determined. In Ref.\cite{gotay}, the authors consider in detail the case    
when $f$ is degenerate, which usually arises when constraints are presented on the system. In which case, $(M,f)$ is called presymplectic manifold. As a consequence, the Hamilton equation, Eq.(\ref{2050a1}), may or may not possess solutions, or possess nonunique solutions. Oppositely, if this matrix is singular and the respective zero-mode does not generate a new constraint, the system has a symmetry. 

The systematization of the symplectic embedding formalism begins by assuming that the gauge invariant version of the general Lagrangian $({\tilde {\cal L}}(a_i,\dot a_i,t))$ is given by
\be
\label{2051}
{\tilde{\cal  L}}(a_i,\dot a_i,\varphi_p,t) = {\cal L}(a_i,\dot a_i,t) + {\cal L}_{WZ}(a_i,\dot a_i,\varphi_p),\,\,\,\,(p = 1,2),
\ee
where $\varphi_p = (\theta, \dot\theta)$ and the extra term $({\cal L}_{WZ})$ depends on the original $(a_i,\dot a_i)$ and WZ $(\varphi_p)$ configuration variables. Indeed, this WZ Lagrangian can be expressed as an expansion in orders of the WZ variable $(\varphi_p)$ such as
\be
\label{wz1}
{\cal L}_{WZ}(a_i,\dot a_i,\varphi_p) = \sum_{n=1}^\infty \upsilon^{(n)}(a_i,\dot a_i, \varphi_p),\;\;{\text with}\;\; \upsilon^{(n)}(\varphi_p)\sim \varphi_p^{n},
\ee
which satisfies the following boundary condition,
\be
\label{wz2}
{\cal L}_{WZ} (\varphi_p=0)= 0.
\ee

The reduction of the Lagrangian, Eq.(\ref{2051}), into its first order form precedes the beginning of conversion process, thus
\be
\label{2052}
{\tilde{\cal L}}^{(0)}dt = A^{(0)}_{\tilde\alpha}d\tilde\xi^{(0){\tilde \alpha}} + \pi_\theta d\theta - {\tilde V}^{(0)}dt,
\ee
where $\pi_\theta$ is the canonical momentum conjugated to the WZ variable, that is,
\be
\label{2052aa2}
\pi_\theta = \frac{\partial{\cal L}_{WZ}}{\partial\dot\theta} = \sum_{n=1}^\infty \frac{\partial\upsilon^{(n)}(a_i, \dot a_i,\varphi_p)}{\partial\dot\theta}.
\ee

The expanded symplectic variables are $\tilde\xi^{(0){\tilde \alpha}} \equiv (a_i, p_i,\varphi_p)$ and the new symplectic potential becomes
\be
\label{2053}
{\tilde V}^{(0)} = V^{(0)} + G(a_i,p_i,\lambda_p),\,\,\,(p = 1,2),
\ee
where $\lambda_p=(\theta,\pi_\theta)$. The arbitrary function $G(a_i,p_i,\lambda_p)$ is expressed as an expansion in terms of the WZ fields, namely
\begin{equation}
\label{2054}
G(a_i,p_i,\lambda_p)= \sum_{n=0}^\infty{\cal G}^{(n)}(a_i,p_i,\lambda_p),
\ee
with 
\be
\label{2054ab}
{\cal G}^{(n)}(a_i,p_i,\lambda_p) \sim \lambda_p^n.
\ee
In this context, the zeroth canonical momenta are given by

\be
{\tilde A}_{\tilde\alpha}^{(0)} = \left\{\begin{array}{lll}
                                  A_{\alpha}^{(0)}, & \mbox{with $\tilde\alpha$ =1,2,\dots,N},\\
                                  \pi_\theta, & \mbox{with ${\tilde\alpha}$= N + 1},\\
                                   0, & \mbox{with ${\tilde\alpha}$= N + 2.}
                                    \end{array}
                                  \right.
\ee
The corresponding symplectic tensor, obtained from the following general relation

\begin{equation}
{\tilde f}_{\tilde\alpha\tilde\beta}^{(0)} = \frac {\partial {\tilde A}_{\tilde\beta}^{(0)}}{\partial \tilde\xi^{(0)\tilde\alpha}} - \frac {\partial {\tilde A}_{\tilde\alpha}^{(0)}}{\partial \tilde\xi^{(0)\tilde\beta}},
\end{equation}
is
\be
\label{2076b}
{\tilde f}_{\tilde\alpha\tilde\beta}^{(0)} = \pmatrix{ { f}_{\alpha\beta}^{(0)} & 0  & 0 
\cr 0 & 0 & - 1
\cr 0 & 1 & 0},
\ee
which should be a singular matrix.

The implementation of the symplectic embedding scheme consists in computing the arbitrary function $(G(a_i,p_i,\lambda_p))$. To this end, the correction terms in order of $\lambda_p$, within by ${\cal G}^{(n)}(a_i,p_i,\lambda_p)$, must be computed as well. If the symplectic matrix, Eq.(\ref{2076b}), is singular, it has a zero-mode $\tilde\varrho$ and, consequently, we have
\begin{equation}
\label{2076}
\tilde\varrho^{(0)\tilde\alpha}{\tilde f}^{(0)}_{\tilde\alpha\tilde\beta} = 0,
\end{equation}
where we assume that this zero-mode is 
\begin{equation}
\label{2076a}
\tilde\varrho^{(0)}=\pmatrix{\gamma^\alpha & 0 & 0},
\end{equation}
where $\gamma^\alpha$, is a generic line matrix. Using the relation given in Eq.(\ref{2076}) together with Eq.(\ref{2076b}) and Eq.(\ref{2076a}), we get
\be
\label{2076c}
\gamma^\alpha{ f}_{\alpha\beta}^{(0)} = 0.
\ee

In this way, a zero-mode is obtained and, in agreement with the symplectic formalism, this zero-mode must be contracted with the gradient of the symplectic potential, namely,
\be
\label{2076c1}
\tilde\varrho^{(0)\tilde\alpha}\frac{\partial \tilde V^{(0)}}{\partial \tilde\xi^{(0)\tilde\alpha}} = 0.
\ee
As a consequence, a constraint arise as being
\be
\label{2076c2}
\Omega = \gamma^\alpha\[\frac{\partial V^{(0)}}{\partial \xi^{(0)\alpha}} + \frac{\partial G(a_i,p_i,\lambda_p)}{\partial \xi^{(0)\alpha}}\].
\ee
Due to this, the first-order Lagrangian is rewritten as
\be
\label{2077}
{\tilde{\cal L}}^{(1)} = A^{(0)}_{\tilde\alpha}\dot{\tilde\xi}^{(0){\tilde \alpha}} + \pi_\theta\dot\theta + \Omega \dot\eta - {\tilde V}^{(1)},
\ee
where ${\tilde V}^{(1)} = V^{(0)}$. Note that the symplectic variables are now $\tilde\xi^{(1)\tilde\alpha}\equiv (a_i,p_i,\eta,\lambda_p)$ (with $\tilde\alpha = 1,2,\dots,N+3$) and the corresponding symplectic matrix becomes
\be
\label{2078}
{\tilde f}_{\tilde\alpha\tilde\beta}^{(1)} = \pmatrix{ { f}_{\alpha\beta}^{(0)} &  {f}_{\alpha\eta} & 0 & 0
\cr {f}_{\eta\beta} & 0 & {f}_{\eta\theta} & {f}_{\eta\pi_\theta}
\cr 0 & {f}_{\theta\eta} & 0 & -1
\cr 0 & {f}_{\pi_\theta\eta} & 1 & 0},
\ee
where
\ba
\label{2078a}
{f}_{\eta\theta} &=& -\frac{\partial}{\partial\theta}\[ \gamma^\alpha\(\frac{\partial V^{(0)}}{\partial \xi^{(0)\alpha}} + \frac{\partial G(a_i,p_i,\lambda_p)}{\partial \xi^{(0)\alpha}}\)\]
,\nonumber\\
{f}_{\eta\pi_\theta} &=& -\frac{\partial}{\partial{\pi_\theta}}\[\gamma^\alpha\( \frac{\partial V^{(0)}}{\partial \xi^{(0)\alpha}} + \frac{\partial G(a_i,p_i,\lambda_p)}{\partial \xi^{(0)\alpha}}\)\],\\
{f}_{\alpha\eta} &=& \frac{\partial \Omega}{\partial \xi^{(0)\alpha}} = \frac{\partial}{\partial \xi^{(0)\alpha}}\[\gamma^\alpha\(\frac{\partial V^{(0)}}{\partial\xi^{(0)\alpha}} + \frac{\partial G(a_i,p_i,\lambda_p)}{\partial\xi^{(0)\alpha}}\)\]
.\nonumber
\ea

Since our goal is to unveil a WZ symmetry, this symplectic tensor must be singular and, consequently, it has a zero-mode, namely,
\be
\label{2078b}
\tilde \nu^{(1)}_{(\nu)(a)} = \pmatrix{\mu^\alpha_{(\nu)} & 1 & a & b},
\ee
which satisfies the relation
\be
\label{2078c}
{\tilde \nu}^{(1)\tilde\alpha}_{(\nu)(a)}{\tilde f}_{\tilde\alpha\tilde\beta}^{(1)}  = 0.
\ee
Note that the parameters $(a,b)$ can be 0 or 1 and $\nu$ indicates the number of choices for ${\tilde \nu}^{(1)\tilde\alpha}$. It is important to notice that $\nu$ is not a fixed parameter. As a consequence, there are two independent set of zero-modes, given by
\ba
\label{2078d}
\tilde \nu^{(1)}_{(\nu)(0)} &=& \pmatrix{\mu^\alpha_{(\nu)} & 1 & 0 & 1},\nonumber\\
\tilde \nu^{(1)}_{(\nu)(1)} &=& \pmatrix{\mu^\alpha_{(\nu)} & 1 & 1 & 0}.
\ea
Note that the matrix elements $\mu^\alpha_{(\nu)}$ present some arbitrariness which can be fixed in order to disclose a desired WZ gauge symmetry. In addition, in our formalism the zero-mode $\tilde\nu^{(1)\tilde\alpha}_{(\nu)(a)}$ is the gauge symmetry generator, which allows to display the symmetry from the geometrical point of view. At this point, we call attention upon the fact that this is an important characteristic since it opens up the possibility to disclose the desired hidden gauge symmetry from the noninvariant model. Different choices of the zero-mode generates different gauge invariant versions of the second class system, however, these gauge invariant descriptions are dynamically equivalent, {\it i.e.}, there is the possibility to relate this set of independent zero-modes, Eq.(\ref{2078d}), through canonical transformation $(\tilde {\bar\nu}^{(\prime,1)}_{(\nu)(a)} = T . \tilde {\bar\nu}^{(1)}_{(\nu)(a)})$ where bar means transpose matrix. For example,
\be
\pmatrix{\mu^\alpha_{(\nu)} \cr 1 \cr 0 \cr 1} = \pmatrix{1 & 0 & 0 & 0\cr 0 & 1 & 0 & 0\cr
0 & 0 & 0 & 1\cr 0 & 0 & 1 & 0} . \pmatrix{\mu^\alpha_{(\nu)} \cr 1 \cr 1 \cr 0}.
\ee
It is important to mention here that, in the context of the BFFT formalism, different choices for the degenerated matrix $X$ leads to different gauge invariant version of the second class model\cite{BN}. Now, it becomes clear that the arbitrariness presents on the BFFT and iterative constraint conversions methods has its origin on the choice of the zero-mode.

From relation Eq.(\ref{2078c}), together with Eq.(\ref{2078}) and Eq.(\ref{2078b}), some differential equations involving  $G(a_i,p_i,\lambda_p)$ are obtained, namely,
\ba
\label{2078e}
0 &=& \mu^\alpha_{(\nu)}{ f}_{\alpha\beta}^{(0)} + { f}_{\eta\beta},\nonumber\\
0 &=& \mu^\alpha_{(\nu)}{ f}_{\alpha\eta}^{(0)} + a { f}_{\theta\eta} + b { f}_{\pi_\theta\eta},\nonumber\\
0 &=& { f}_{\eta\theta}^{(0)} + b,\\
0 &=& { f}_{\eta\pi_\theta}^{(0)} - a.\nonumber
\ea
Solving the relations above, some correction terms, within $\sum_{m=0}^\infty {\cal G}^{(m)}(a_i,p_i,\lambda_p)$, can be determined, also including the boundary conditions $({\cal G}^{(0)}(a_i, p_i,\lambda_p = 0 ))$.

In order to compute the remaining corrections terms of $G(a_i,p_i,\lambda_p)$, we impose that no more constraints arise from the contraction of the zero-mode $(\tilde\nu^{(1)\tilde\alpha}_{(\nu)(a)})$ with the gradient of potential ${\tilde V}^{(1)}(a_i,p_i,\lambda_p)$. This condition generates a general differential equation, which reads as
\begin{eqnarray}
\label{2080}
0 &=& \tilde\nu^{(1)\tilde\alpha}_{(\nu)(a)}\frac{\partial {\tilde  V}^{(1)}(a_i,p_i,\lambda_p)}{\partial{\tilde\xi}^{(1)\tilde\alpha}}
\nonumber\\
&=& \mu^\alpha_{(\nu)} \left[\frac{\partial {V}^{(1)}(a_i,p_i)}{\partial{\xi}^{(1)\alpha}} + \frac{\partial  G(a_i,p_i,\theta,\pi_\theta)}{\partial{\xi}^{(1)\alpha}}\right] + a \frac{\partial G(a_i,p_i,\lambda_p )}{\partial\theta} + b\frac{\partial G(a_i,p_i,\lambda_p)}{\partial\pi_\theta}\nonumber\\
&=& \mu^\alpha_{(\nu)} \left[\frac{\partial {V}^{(1)}(a_i,p_i)}{\partial{\xi}^{(1)\alpha}} + \sum_{m=0}^\infty\frac{\partial {{\cal G}}^{(m)}(a_i,p_i,\lambda_p )}{\partial{\xi}^{(1)\alpha}}\right] + a \sum_{n=0}^\infty\frac{\partial {\cal G}^{(n)}(a_i,p_i,\lambda_p )}{\partial\theta}\nonumber\\
&+& b \sum_{m=0}^\infty\frac{\partial {{\cal G}}^{(n)}(a_i,p_i,\lambda_p )}{\partial\pi_\theta}.
\end{eqnarray}
The last relation allows us to compute all correction terms in order of $\lambda_p$, within ${\cal G}^{(n)}(a_i,p_i,\lambda_p)$. Note that this polynomial expansion in terms of $\lambda_p$ is equal to zero, subsequently, all the coefficients for each order in this WZ variables must be identically null. In view of this, each correction term in orders of $\lambda_p$ can be determined as well. For a linear correction term, we have
\be
\label{2090}
0 = \mu^\alpha_{(\nu)} \left[\frac{\partial {V}^{(0)}(a_i,p_i)}{\partial{\xi}^{(1)\alpha}} + \frac{\partial {{\cal G}}^{(0)}(a_i,p_i)}{\partial{\xi}^{(1)\alpha}}\right] + a \frac{\partial{\cal G}^{(1)}(a_i,p_i,\lambda_p)}{\partial\theta} + b\frac{\partial{{\cal G}}^{(1)}(a_i,p_i,\lambda_p)}{\partial\pi_\theta},
\ee
where the relation $V^{(1)} = V^{(0)}$ was used. For a quadratic correction term, we get
\be
\label{2095}
0 = {\mu}^{\alpha}_{(\nu)}\left[\frac{\partial{\cal G}^{(1)}(a_i,p_i,\lambda_p)}{\partial{\xi}^{(0)\alpha}} \right] + a \frac{\partial{\cal G}^{(2)}(a_i,p_i,\lambda_p)}{\partial\theta}  + b \frac{\partial{{\cal G}}^{(2)}(a_i,p_i,\lambda_p)}{\partial\pi_\theta}.
\ee
From these equations, a recursive equation for $n\geq 2$ is proposed as

\be
\label{2100}
0 = {\mu}^{\alpha}_{(\nu)}\left[\frac{\partial {\cal G}^{(n - 1)}(a_i,p_i,\lambda_p)}{\partial{\xi}^{(0)\alpha}} \right] + a\frac{\partial{\cal G}^{(n)}(a_i,p_i,\lambda_p)}{\partial\theta} + b \frac{\partial{{\cal G}}^{(n)}(a_i,p_i,\lambda_p)}{\partial\pi_\theta},
\ee
which allows us to compute the remaining correction terms in order of $\theta$ and $\pi_\theta$. This iterative process is successively repeated up to Eq.(\ref{2080}) when it becomes identically null or when an extra term ${\cal G}^{(n)}(a_i,p_i,\lambda_p)$ can not be computed. Then, the new symplectic potential is written as
\be
\label{2110}
{\tilde  V}^{(1)}(a_i,p_i,\lambda_p) = V^{(0)}(a_i,p_i) + G(a_i,p_i,\lambda_p).
\end{equation}
For the  case {\it i}, the new symplectic potential is gauge invariant. For the second case  {\it ii}, due to some corrections terms within $G(a_i,p_i,\lambda_p)$ that are not yet determined, this new symplectic potential is not gauge invariant. As a consequence, there are some WZ counter-terms in the new symplectic potential,which can be fixed using Hamilton's equation of motion for the WZ variables $\theta$ and $\pi_\theta$ together with the canonical momentum relation conjugated to $\theta$, given in Eq.(\ref{2052aa2}). Due to this, the gauge invariant Hamiltonian is obtained explicitly and the zero-mode ${\tilde\nu}^{(1)\tilde\alpha}_{(\nu)(a)}$ is identified as being the generator of the infinitesimal gauge transformation, given by

\begin{equation}
\label{2120}
\delta{\tilde\xi}^{\tilde\alpha}_{(\nu)(a)} = \varepsilon{\tilde\nu}^{(1)\tilde\alpha}_{(\nu)(a)},
\end{equation}
where $\varepsilon$ is an infinitesimal parameter.

\section{Symplectic Quantization of the Noncommutative Massive $U(1)$ Theory}
\label{sec:III}

The lagrangian density that govern the dynamics of noncommutative massive $U(1)$ theory is
 
\begin{equation}
\label{equation1}
{\cal L} = - \,\,\frac{1}{4}F_{\mu\nu}F^{\mu\nu} + \frac{1}{2}\,m^2\,A^{\mu}A_{\mu},
\end{equation}
where the stress tensor in terms of Moyal commutator is given by 

\begin{equation}
\label{equation2}
F_{\mu\nu} = \partial_{\mu}A_{\nu} - \partial_{\nu}A_{\mu} 
- ie\[A_\mu,A_\nu\],
\end{equation}
with

\begin{equation}
\label{equation3}
\[A_\mu,A_\nu\] = A_{\mu} \star A_{\nu} - A_{\nu} \star A_{\mu},
\end{equation}
and where

\begin{eqnarray}
\label{equation4}
A_{\mu}(x) \star A_{\nu}(x) &=& exp(\frac{i}{2}\theta^{\gamma\lambda}
\partial_{\gamma}^{x}\partial_{\lambda}^{y})
A_{\mu}(x)A_{\nu}(y)\big\vert_{x=y},\nonumber \\
A_{\nu}(x) \star A_{\mu}(x) &=& exp(\frac{i}{2}\theta^{\lambda\gamma}
\partial_{\lambda}^{x}\partial_{\gamma}^{y})
A_{\nu}(x)A_{\mu}(y)\big\vert_{x=y},
\end{eqnarray}
where $\theta^{\gamma\lambda}$ is a real and antisymmetric constant matrix. In order to avoid causality and unitary problems in the Moyal space, we take 
$\theta^{0i} = 0$\cite{Gomis}. Hence the $\star$ product of the gauge fields into the stress tensor, given in Eq.(\ref{equation2}), becomes

\begin{eqnarray}
\label{equation5}
A_{\mu}(x) \star A_{\nu}(x) &=& exp(\frac{i}{2}\theta^{ij}
\partial_{i}^{x}\partial_{j}^{y})
A_{\mu}(x)A_{\nu}(y)\big\vert_{x=y},\nonumber \nonumber\\
A_{\nu}(x) \star A_{\mu}(x) &=& exp(\frac{i}{2}\theta^{ji}
\partial_{j}^{x}\partial_{i}^{y})
A_{\nu}(x)A_{\mu}(y)\big\vert_{x=y}.
\end{eqnarray}

Now, we are ready to reduce the second-order lagrangian density, Eq.((\ref{equation1})), into its first-order form, which is read as
 
\begin{eqnarray}
\label{equation6}
{\cal L} &=& \pi^i\dot A_i + A_0(\partial_i\pi^i + m^2A^0) 
+ \frac{1}{2}\pi_i\pi^i \nonumber \nonumber\\ 
&-& ie\pi^i(A_0 \star A_i - A_i \star A_0)
- \frac{1}{4}F_{ij}F^{ij} + \frac{1}{2}m^2A_iA^i \nonumber\\
&-& \frac{1}{2}m^2A_0A^0,
\end{eqnarray}
with the canonical momentum $\pi_i$ given by

\begin{eqnarray}
\label{equation7}
\pi_i &=& - F_{0i}\nonumber\\
&=& -\dot A_i + \partial_iA_0 + ie(A_0 \star A_i - A_i \star A_0).
\end{eqnarray}

The symplectic fields are $\xi^{(0)\alpha}=(A^i,\pi^i,A^0)$ and the zeroth-iterative symplectic matrix is

\begin{equation}
\label{equation8}
f^{(0)} = \left(
\begin{array}{ccc}
0           & -\delta^{i}_{j} & 0 \\
\delta^{j}_{i}&         0     & 0 \\
0           &         0     & 0
\end{array}
\right)\,\delta({x} - {y}),
\end{equation}
which is a singular matrix. It has a zero-mode that generates the following constraint

\begin{equation}
\label{equation9}
\Omega(x) = \partial^{x}_{i}\pi^i(x) + m^2A^0(x) - ie\[A_i(x),\pi^i(x)\],
\end{equation}
identified as being the Gauss law. Bringing back this constraint into the canonical part of the first-order lagrangian density ${\cal L}^{(0)}$ using a lagrangian multiplier ($\beta$), the first-iterated lagrangian density, written in terms of $\xi^{(1)\alpha} = (A^i,\pi^i,A^0, \beta)$ is obtained as

\begin{eqnarray}
\label{equation10}
{\cal L}^{(1)} &=& \pi^{i}\dot A_i + \dot{\beta}\Omega + \frac{1}{2}\pi_i\pi^i - \frac{1}{4}F_{ij}F^{ij} \nonumber\\
&+& \frac{1}{2}m^2A_iA^i - \frac{1}{2}m^2A_0A^0, 
\end{eqnarray}
with the following symplectic fields $\xi^{(1)\alpha}=(A^i,\pi^i,A^0,\beta)$.
The first-iterated symplectic matrix is obtained as being

\begin{equation}
\label{equation11}
f^{(1)}=\left(
\begin{array}{cccc}
0           & -\delta^i_{j}\delta(x - y)  &  0   &   ie\[\pi^i(y),\delta(x - y)\]\\
\delta^j_{i}\delta(x - y)  &         0     &   0    &   f_{\pi_i\beta}  \\
0         &         0     &   0    &     m^2\delta(x - y)   \\
ie\[\delta(x - y),\pi^i(x)\]        &        f_{\beta\pi_i}    &  -m^2 \delta(x - y)     &     0
\end{array}
\right).
\end{equation}
where

\be
\label{D1}
f_{\pi_i\beta}(x,y) = \partial^y_i\delta(x - y) + ie\[\delta(x - y),A^i(y)\].
\ee

This matrix is nonsingular and, as settle by the symplectic formalism, the Dirac brackets among the phase space fields are acquired from the inverse of the symplectic matrix, namely,

\ba
\label{equation12}
\lbrace A_i(x),A^j(y)\rbrace^* &=& 0,\nonumber\\
\lbrace A_i(x),\pi^j(y)\rbrace^* &=& \delta_i^j\delta(x - y),\nonumber\\
\lbrace A_i(x),A_0(y)\rbrace^* &=& - \frac{1}{m^2}\(\partial^x_i\delta(x - y) + \frac{ie}{m^2}\[\delta(x - y),A_i(y)\]\),\nonumber\\
\lbrace \pi^i(x),A_0(y)\rbrace^* &=& \frac{ie}{m^2}\[\delta(x - y),\pi^i(y)\].
\ea

Afterwards, we are ready to implement the symplectic embedding formalism of the theory. This will be done in the next section.

\section{The embedded model}
\label{sec:IV}

At this point we are interested to embed the massive noncommutative $U(1)$ theory {\it via} symplectic embedding formalism (Sec. \ref{sec:II}) \cite{ANO}. The symplectic embedding process begins enlarging the phase space with the introduction of two WZ fields $\gamma = \pmatrix{\eta&\pi_\eta}$. Due to this, the original Lagrangian density, Eq.(\ref{equation1}), becomes

\begin{equation}
\label{equation1a}
{\tilde{\cal L}} = {\cal L} + {\cal L}_{WZ},
\end{equation}
where ${\cal L}_{WZ}$ is a WZ counter-term which eliminates the noncommutativity of the theory. In agreement with symplectic embedding formalism, this new Lagrangian density must be reduced into its first-order form, namely,

\begin{eqnarray}
\label{equation13}
{\tilde{\cal L}}^{(0)} &=& \pi^i\dot A_i + \pi_\eta \dot \eta - {\tilde V}^{(0)},
\end{eqnarray}
where ${\tilde V}^{(0)}$ is the symplectic potential , given by 
 
\begin{eqnarray}
\label{equation14}
{\tilde V}^{(0)} &=& - A_0(\partial_i\pi^i + m^2A^0) 
- \frac{1}{2}\pi_i\pi^i \nonumber \nonumber\\ 
&+& ie\pi^i(A_0 \star A_i + A_i \star A_0)
+ \frac{1}{4}F_{ij}F^{ij} - \frac{1}{2}m^2A_iA^i \nonumber\\
&+& \frac{1}{2}m^2A_0A^0 + G(A^i,\pi^i,A^0,\gamma),
\end{eqnarray}
where $G\equiv G(A^i,\pi^i,A^0,\gamma)$ is a arbitrary function and is written as an expansion in terms of the WZ fields as

\begin{equation}
G(A^i,\pi^i,A^0,\gamma) = \sum_{n=0}^\infty {\cal G}^{(n)}(A^i,\pi^i,A^0,\gamma)\,\,\,\text{with}\,\,\,\, {\cal G}^{(n)}(A^i,\pi^i,A^0,\gamma)\sim (\gamma)^{n}.
\end{equation}

The new symplectic variables are now given by ${\tilde\nu}^{(0)\alpha}=(A^i,\pi^i,A^0,\gamma)$ and the respective symplectic tensor is

\be
\label{matrix00}
{\tilde f}^{(0)} = \pmatrix{ 0 & - \delta_j^i\delta(x - y) & 0 & 0 & 0
\cr \delta_i^j\delta(x - y) &  0 & 0 & 0 & 0
\cr 0 & 0 & 0 & 0 & 0
\cr 0 & 0 & 0 & 0 & - \delta (x - y)
\cr 0 & 0 & 0 & \delta (x - y) & 0}.
\ee
This singular matrix has a zero-mode, which is settle as

\begin{equation}
{\tilde\nu}^{(0)} = \pmatrix{0 & 0 & 1 & 0 & 0}.
\end{equation}
This zero-mode when contracted with the symplectic potential generates the following constraint,

\begin{equation}
\Omega (x) = \partial^x_i\pi^i(x) + m^2A^0(x) - i e[A^i(x),\pi_i(x)] + \int dy \frac{\delta G(y)}{\delta A^0(x)}.
\end{equation}

In accordance with the symplectic scheme, this constraint must be introduced into the zeroth-iterative first-order Lagrangian density through a Lagrange multiplier $\zeta$, generating the next one,

\begin{equation}
{\tilde{\cal L}}^{(1)} = \pi^i\dot A_i + \pi\dot \eta + \Omega\dot\zeta - {\tilde V}^{(1)},
\end{equation}
with ${\tilde V}^{(1)}={\tilde V}^{(0)}\mid_{\Omega=0}$. Now, the symplectic vector is ${\tilde\xi}^{(1)} = (A^i,\pi^i,A^0,\zeta,\gamma)$ with the corresponding tensor given by

\begin{equation}
{\tilde f}^{(1)} = \pmatrix{ 0 & - \delta_i^j\delta(x - y) & 0 & \frac{\delta\Omega(y)}{\delta A^i(x)} & 0 & 0
\cr \delta_j^i\delta(x - y) &  0 & 0 & \frac{\delta\Omega(y)}{\delta \pi^i(x)} & 0 & 0
\cr 0 & 0 & 0 & \frac{\delta\Omega(y)}{\delta A^0(x)} & 0 & 0
\cr - \frac{\delta\Omega(x)}{\delta A^j(y)} & - \frac{\delta\Omega(x)}{\delta \pi^j(y)} & - \frac{\delta\Omega(x)}{\delta A^0(y)} & 0 & - \frac{\delta\Omega(x)}{\delta \eta(y)} & - \frac{\delta\Omega(x)}{\delta \pi_\eta(y)}
\cr 0 & 0 & 0 &   \frac{\delta\Omega(y)}{\delta \eta(x)} & 0 & - \delta(x - y)
\cr 0 & 0 & 0 &  \frac{\delta\Omega(y)}{\delta \pi_\eta(x)}  & \delta(x - y) & 0}.
\ee

Now, we are ready to remove the noncommutative character of the original theory. To this end, it is necessary to assume that the symplectic matrix above is singular. Consequently, this matrix has a respective zero-mode, which is degenerated due to  the arbitrariness present on the matrix, which resides on the degenerated function $G$. This is not bad at all since it gives room to settle several approaches to eliminate the noncommutative structure. Consequently, this opens up the possibility to obtain several commutative embedded representations for the noncommutative model, whose all of them are dynamically equivalent. This represents a quite feature of the symplectic embedding formalism. In view of this, we chose a convenient zero-mode, which is write as

\begin{equation}
\label{equation15}
{\tilde \nu}^{(1)} = \pmatrix{\partial^{x,i} & 0 & 0 & -1 & 1 & 1}.
\end{equation}

Contracting this zero-mode with the symplectic matrix above, namely,

\begin{equation}
\label{wo1}
\int d^3 x\,\, \nu^{(1) \tilde \alpha}(x) {\tilde f}^{(1)}_{\tilde \alpha \tilde \beta}(x,y) = 0,
\end{equation}
we get the boundary condition ${\cal G}^{(0)}$ as
 
\be
\label{matrix04}
{\cal G}^{(0)}(x) = - \frac 12 m^2 A_0(x)A^0(x) + ie\[A^i(x),\pi_i(x)\] A^0(x),
\ee
and some of the correction terms belong to ${\cal G}^{(1)},$ namely, $\pi_\eta A^0 - \eta A^0$. We note that the correction term ${\cal G}^{(n)}$ for $n\geq 2$ has no dependence on the temporal component of potential field $A^0$.Thus, ${\cal G}^{(n)}\equiv{\cal G}^{(n)}(A^i,\pi^i,\gamma)$ for $n\geq 2$. This completes the first step of the symplectic embedding formalism.

After introducing these correction terms into the symplectic potential ${\tilde V}^{(1)}$, let us begin with the second step in order to reformulate the theory as a gauge theory. Following the prescription of the symplectic embedding formalism, the zero-mode ${\tilde \nu}^{(1)}$ does not produce a constraint when contracted with the gradient of symplectic potential, namely,

\be
\label{matrix06}
\int d^3x \;\;\tilde\nu^{(1)\tilde\alpha}(y)\;\frac{\delta {\tilde V}^{(1)}(x)}{\delta {\tilde\xi}^{\tilde\alpha}(y)} = 0,
\ee
instead of, it produces a general equation that allows the computation of the correction terms in order of $\gamma$ enclosed into $G(A_i,\pi_i,A_0,\gamma)$. To compute the remains linear correction terms in $\gamma$, ${\cal G}^{(1)}$, we use the following relation ({\it vide} \ref{2090})
 
\begin{eqnarray}
\label{matrix07}
0 &=& \int\; d^3x\; \,\{- ie\[F_{ij}(x),\,A^i(x)\]\partial^{j}_y\delta(x-y)\nonumber\\ 
&-&  m^2A_j(x)\partial^{j}_y\delta(x-y) + ie\[\pi_i(x),\,A_0(x)\]\partial^{i}_y\delta(x-y)\nonumber\\
&+& \frac{\delta{\cal G}^{(1)}(x)}{\delta\eta(y)} + \frac{\delta{\cal G}^{(1)}(x)}{\delta\pi_\eta(y)}\}.
\end{eqnarray}
After a straightforward calculation, the complete linear correction term $\gamma$ is given by 

\begin{eqnarray}
\label{matrix08}
{\cal G}^{(1)}(x)  &=& \pi_\eta(x) A^0(x) - \eta(x) A^0(x) + \{ie\partial^{j}_x\[F_{ij}(x),\,A^i(x)\]\nonumber\\
&+& ie\partial^{j}_x \[\pi_j(x),\,A_0(x)\]
 + m^2\partial^{j}_xA_j(x)\}\frac 12 (\eta(x) + \pi_\eta(x)).
\end{eqnarray}

In order to compute the quadratic correction term, we use the following relation ({\it vide} \ref{2095})

\be
\label{matrix09}
\int d^3x\,\, \[\partial^{j}_x \(\frac {\delta{\cal G}^{(1)}}{\delta A^j(x)} \) + \frac{\delta{\cal G}^{(2)}(y)}{\delta\eta(x)} + \frac{\delta{\cal G}^{(2)}(y)}{\delta\pi_\eta(x)}\] = 0,
\ee
which after a direct calculation, we get

\begin{eqnarray}
\label{matrix10}
{\cal G}^{(2)}(x)  &=& - \frac{ie}{4} F_{i,j}\[\partial^{i}_x\eta(x),\,\partial^{j}_x\eta(x)\] - \frac 14 m^2\,\, \,\partial^j_{x}\eta(x)\partial_j^{x}\eta(x)\nonumber \\
&-& \frac{e^2}{4} \[\partial^{i}_x\eta(x),\,A_j(x)\] \[A_i(x),\,\partial^{j}_x\eta(x)\] \nonumber \\
&-&\frac{e^2}{4}\[A_i(x),\,\partial_j^{x}\eta(x)\]
\[A^i(x),\,\partial^{j}_x\eta(x)\]\nonumber\\
&-& \frac{ie}{4} F_{i,j}\[\partial^{i}_x\pi_\eta(x),\,\partial^{j}_x\pi_\eta(x)\] - \frac 14 m^2\,\, \,\partial^j_{x}\pi_\eta(x)\partial_j^{x}\pi_\eta(x)\nonumber \\
&-& \frac{e^2}{4} \[\partial^{i}_x\pi_\eta(x),\,A_j(x)\] \[A_i(x),\,\partial^{j}_x\pi_\eta(x)\] \nonumber \\
&-&\frac{e^2}{4}\[A_i(x),\,\partial_j^{x}\pi_\eta(x)\]
\[A^i(x),\,\partial^{j}_x\pi_\eta(x)\].
\end{eqnarray}

The remaining corrections ${\cal G}^{(n)}$ for $n\geq 3$ are computed in analogous way ({\it vide} \ref{2100}) and we just write them down as

\begin{eqnarray}
\label{matrix11}
{\cal G}^{(3)}(x) &=& \frac{e^2}{2} \[A_i(x),\,\partial_j^{x}\eta(x)\]
\[\partial^{j}_x\eta(x),\,\partial^{y,i}\eta(x)\],\nonumber \\
&+& \frac{e^2}{2}\[A_i(x),\,\partial_j^{x}\pi_\eta(x)\]
\[\partial^{j}_x\pi_\eta(x),\,\partial^{y,i}\pi_\eta(x)\],\nonumber \\
{\cal G}^{(4)}(x) &=& \frac{e^2}{8} \[\partial_i^{x}\eta(x),\,\partial_j^{x}\eta(x)\]
\[\partial^{j}_x\eta(x),\,\partial^{i}_x\eta(x)\]\nonumber\\
&+& \frac{e^2}{8} \[\partial_i^{x}\pi_\eta(x),\,\partial_j^{x}\pi_\eta(x)\]
\[\partial^{j}_x\pi_\eta(x),\,\partial^{i}_x\pi_\eta(x)\]. 
\end{eqnarray}

Note that the fourth-order correction term has dependence on the WZ field only, thus all correction terms ${\cal G}^{(n)}$ for $n\geq 5$ are null. Then, the gauge invariant first-order Lagrangian density is given by 

\begin{equation}
{\tilde{\cal L}} = \pi^i(x)\dot A_i(x) + \pi(x)\dot \eta(x) - {\tilde {\cal H}},
\end{equation}
where ${\tilde {\cal H}}$ is the gauge invariant Hamiltonian density, identified as being the symplectic potential ${\tilde V}^{(1)}$, namely,

\begin{eqnarray}
{\tilde {\cal H}} &=& {\tilde V}^{(1)} = - \frac{1}{2}\pi_i(x)\pi^i(x) + \frac{1}{4}F_{ij}F^{ij} - \frac{1}{2}m^2A_i(x)A^i(x) - ie\[A^i(x), A_0(x)\]\pi_i(x) \nonumber\\
&+& 2\pi_\eta(x) A^0(x) - 2\eta(x) A^0(x)  
+ \frac{ie}{2}\partial^{j}_x\[F_{ij}(x),\,A^i(x)\]\eta(x) + \frac 12 m^2\partial^{j}_xA_j(x)\eta(x) \nonumber\\
&-& \frac{ie}{2}\partial^{j}_x\[\pi_i, A_0\]\eta(x) 
- \frac{ie}{4} F_{i,j}\[\partial^{i}_x\eta(x),\,\partial^{j}_x\eta(x)\] \nonumber \\
&-& \frac 14 m^2\,\, \,\partial^j_{x}\eta(x)\partial_j^{x}\eta(x) - \frac{e^2}{4} \[\partial^{i}_x\eta(x),\,A_j(x)\] \[A_i(x),\,\partial^{j}_x\eta(x)\] \nonumber \\
&-&\frac{e^2}{4}\[A_i(x),\,\partial_j^{x}\eta(x)\]
\[A^i(x),\,\partial^{j}_x\eta(x)\] + \frac{e^2}{2} \[A_i(x),\,\partial_j^{x}\eta(x)\]
\[\partial^{j}_x\eta(x),\,\partial^{y,i}\eta(x)\]\nonumber\\
&+& \frac{e^2}{8} \[\partial_i^{x}\eta(x),\,\partial_j^{x}\eta(x)\]
\[\partial^{j}_x\eta(x),\,\partial^{i}_x\eta(x)\]\nonumber\\
&+& \frac{ie}{2}\partial^{j}_x\[F_{ij}(x),\,A^i(x)\]\pi_\eta(x) + \frac 12 m^2\partial^{j}_xA_j(x)\pi_\eta(x) \nonumber\\
&-& \frac{ie}{2}\partial^{j}_x\[\pi_i, A_0\]\pi_\eta(x) - \frac{ie}{4} F_{i,j}\[\partial^{i}_x\pi_\eta(x),\,\partial^{j}_x\pi_\eta(x)\] \nonumber \\
&-& \frac 14 m^2\,\, \,\partial^j_{x}\pi_\eta(x)\partial_j^{x}\pi_\eta(x) - \frac{e^2}{4} \[\partial^{i}_x\pi_\eta(x),\,A_j(x)\] \[A_i(x),\,\partial^{j}_x\pi_\eta(x)\] \nonumber \\
&-&\frac{e^2}{4}\[A_i(x),\,\partial_j^{x}\pi_\eta(x)\]
\[A^i(x),\,\partial^{j}_x\pi_\eta(x)\] \nonumber\\
&+& \frac{e^2}{2}\[A_i(x),\,\partial_j^{x}\pi_\eta(x)\]
\[\partial^{j}_x\pi_\eta(x),\,\partial^{y,i}\pi_\eta(x)\] \nonumber\\
&+& \frac{e^2}{8} \[\partial_i^{x}\pi_\eta(x),\,\partial_j^{x}\pi_\eta(x)\]
\[\partial^{j}_x\pi_\eta(x),\,\partial^{i}_x\pi_\eta(x)\].
\end{eqnarray}

In order to complete the gauge invariant reformulation of the massive noncommutative $U(1)$ theory, we compute the infinitesimal gauge transformations of the phase space coordinates. In agreement with the symplectic method, the zero-mode ${\tilde \nu}^{(1)}$ is the generator of the infinitesimal gauge transformations $(\delta {\cal O} = \varepsilon {\tilde \nu}^{(1)})$, which are given by
\begin{eqnarray}
\label{equation16}
\delta A_i &=& \partial^i \varepsilon,\nonumber\\
\delta \pi^i &=& 0,\nonumber\\
\delta A_0 &=& 0,\\
\delta\eta &=& \varepsilon,\nonumber\\
\delta\pi_\eta &=& \varepsilon,\nonumber\\
\end{eqnarray}
where $\varepsilon(y)$ is an infinitesimal time-dependent parameter.

Thus, we complete the Hamiltonian symplectic embedding of the massive noncommutative $U(1)$ theory. 

\section{Conclusion}
\label{sec:V}

In this paper, we have embedded the massive noncommutative $U(1)$ theory. This was achieved through an alternative embedding formalism, based on the symplectic framework. The  Hamiltonian density of the embedded version of the massive noncommutative $U(1)$ theory was also obtained. A remarkable feature gift in the present work is that the embedded version was obtained after a finite number of steps in the iterative symplectic embedding process, which leads to a embedded Hamiltonian density with a finite number of WZ terms. Further, we also discuss, in Section \ref{sec:III}, the new possibilities to embedded the noncommutative theory. It is important to regard that, by construction, these different embedded representations of the noncommutative theory are dynamically equivalent, since the WZ gauge orbit is defined by the zero-mode. In view of this, the symplectic embedding formalism  seems powerful enough when compared with other WZ conversion schemes.

\section{ Acknowledgments}
This work was supported in part by Funda\c c\~ao de Amparo a Pesquisa do Estado de Minas Gerais (FAPEMIG) and Conselho Nacional de Desenvolvimento Cient\'\i fico e Tecnol\'ogico (CNPq), Brazilian Research Agencies.

\appendix

\section{Some properties of the Moyal product}

In this appendix we list some properties that we use in this paper.

\be
\label{woa}
\int d^nx\,\,\phi_1 \star \phi_2 = \int d^nx\,\,\phi_1\phi_2 = \int d^nx\,\,\phi_2 \star \phi_1,
\ee
\be
\label{wob}
(\phi_1\star\phi_2)\star\phi_3 = \phi_1\star(\phi_2\star\phi_3) = \phi_1\star\phi_2\star\phi_3,
\ee
\ba
\label{woc}
\int d^nx\,\,\phi_1\star\phi_2\star\phi_3 &=& 
\int d^nx\,\,\phi_2\star\phi_3\star\phi_1\nonumber\\
&=& \int d^nx\,\,\phi_3\star\phi_1\star\phi_2,
\ea  
\ba
\label{wod}
\int d^nx\,\,\[\[A,B\],C\]\star D &=& \int d^nx\,\,\[A,B\]\star \[C,D\]\nonumber\\
&=& \int d^nx\,\,\[A,B\] \[C,D\].
\ea

\end{document}